\documentclass[pre,eqsecnum,showpacs,byrevtex,amsmath,amssymb,nofootinbib]{revtex4}

 \usepackage{amsmath,graphicx}

\begin{document}

\title{Localization and freezing of a Gaussian chain in a quenched
  random potential}
 
\author{ Vakhtang G. Rostiashvili and 
   Thomas A.  Vilgis}
 \affiliation{Max Planck Institute for Polymer Research\\
   10 Ackermannweg, 55128 Mainz, Germany.}

\begin{abstract}
  The Gaussian chain in a quenched random potential (which is characterized by
  the disorder strength $\Delta$) is investigated in the $d$ - dimensional
  space by the replicated variational method. The general expression for the
  free energy within so called one - step - replica symmetry breaking (1 -
  RSB) scenario has been systematically derived. We have shown that the
  replica symmetrical (RS) limit of this expression can describe the chain
  center of mass localization and collapse. The critical disorder when the
  chain becomes localized scales as $\Delta_c \simeq b^d N^{-2 + d/2}$ ( where
  $b$ is the length of the Kuhn segment length and $N$ is the chain length)
  whereas the chain gyration radius $R_{\rm g} \simeq b \left(b^d/\Delta
  \right)^{1/(4 - d)}$. The freezing of the internal degrees of freedom
  follows to the 1-RSB - scenario and is characterized by the beads
  localization length $\overline{{\cal D}^2}$. It was demonstrated that the
  solution for $\overline{{\cal D}^2}$ appears as a metastable state at
  $\Delta = \Delta_A$ and behaves similarly to the corresponding frozen states
  in heteropolymers or in $p$ - spin random spherical model.
\end{abstract}
\maketitle

\section{Introduction}
The behavior of polymer chains (with or without excluded volume) in a quenched
random potential is a simple but nontrivial problem of statistical mechanics
with disorder \cite{Baum}. Despite the vast investigations of both statical
\cite{Edwards,Cates,Machta,Machta1,Natter,Obukh,Step,Vilgis1,Gold1,Gold2} and
dynamical \cite{Step1,Ebert,Ebert1} properties many aspects remain to be
elucidate. First of all we emphasize that one should discriminate between the
annealed and quenched problems. In the case that the effective diffusion
coefficient is large enough, so that the chain experiences different disorder
realizations, the problem is equivalent to the annealed one (even though the
random potential is quenched). In this case the presence of the disorder
manifests itself (in the statics) through the reduction of the excluded volume
parameter $v$, i.e. $v \to v - \Delta$, where $\Delta$ is the second moment of
the quenched random potential. Strictly speaking it was shown long time ago
\cite{Cates}, mainly by the use of Imry - Ma arguments, that in the infinite
volume limit the quenched and annealed problems coincide, so that the Gaussian
chain collapses to a size of one monomer. Only if the volume of the random
medium $\Omega$ is finite the chain acquires the characteristic gyration
radius $R_{\rm g} \approx b[(\Delta/b^b)\ln (\Omega/b^d)]^{-1/(4-d)}$, where
$b$ is the Kuhn segment length. Nevertheless this approach completely ignores
the very important dynamic aspects of the problem.

Indeed, as the disorder strength $\Delta$ and the chain length $N$ increase,
the diffusion coefficient $D$ falls down dramatically and the chain can not
sample the whole random medium over the course of internal degrees of freedom
equilibration. It was shown \cite{Machta,Machta1} that in the $d$-dimensional
space the center of mass diffusion coefficient $D \approx D_{\rm R} \exp[-
(\Delta/b^d)N^{\alpha}]$, where $D_{\rm R}$ is the Rouse diffusion coefficient,
$b$ is the Kuhn segment length, $\alpha = 2 - \nu d$ and $\nu$ is the Flory
exponent. As a result the time needed to obtain an ``averaged'' (annealed)
statistics grows exponentially with $\Delta$ and $N$. In a recent publication
\cite{Miglior} we have treated this problem dynamically by functional tools
based on a Langevin dynamics approach. This approach avoids replicas naturally
but by averaging over the quenched disorder inevitably couples different
dynamic trajectories. This leads ultimately to the non - Markovian diffusional
slowing down as well as to the corresponding freezing and non - ergodic
regimes for the chain Rouse modes.  We have derived the equation of motion for
the Rouse mode time - dependent correlation function within a Hartree
approximation and found that this equation has a memory kernel which is generic
for the mode - coupling theory of the glass transition \cite{Gotz}.  The self
- consistent treatment of the mode - coupling equations leads to the
conclusion that the chain center of mass diffusion coefficient decreases
according to the law: $D \approx D_{\rm R}[1 - {\rm const}
(\Delta/b^d)N^{\alpha}]$. The individual Rouse modes of the chain also freeze
at a common disorder strength, but their dynamic pathway is strongly selective
to the mode number.  We have recently launched the intensive Monte - Carlo
(MC) simulation \cite{Milchev}  to support these analytical calculations.
Indeed we have found that the diffusion coefficient as a function of
$(\Delta/b^d)N^{\alpha}$ drops much more dramatic than the exponential law,
which is mainly based on the simple Markovian diffusion approximations.
Preliminary MC - results \cite{Milchev} can be fitted much better by the form
$D \sim [1 - {\rm const} (\Delta/b^d)N^{\alpha}]$. Thus suggests that a
critical disorder, $\Delta_{\rm c} \approx b^d N^{- \alpha}$,exists  and that for
strong disorder $\Delta > \Delta_{\rm c}$ the chain center of mass is
localized under the conditions of the simulation.
  
Such a dramatic reduction of the diffusion coefficient shows that
because the
dynamics of the  chain becomes indeed extremely slow the equivalence of the
quenched and annealed problems in the infinite volume limit
\cite{Cates,Natter,Gold1,Gold2} has to be questioned for strong disorder,
i.e., $\Delta > \Delta_{\rm c}$. As soon as $\Delta > \Delta_{\rm c}$ the
chain center of mass is practically localized and the chain may only sample
intermediate disorder environment. We should implement in this case the
quenched averaging,i.e. the chain free energy (or any observable physical
quantity) must be calculated first for a particular configuration of disorder,
only after this the average over the disorder can be taken. In any case, as
proposed by the ref.\cite{Cates,Natter,Gold1,Gold2}, the system volume
indicated by the factor $\ln \Omega$ represents then the actually explored
volume.  For completeness we mention that the MC - simulations \cite{Milchev}
also reveal the Rouse modes freezing by showing a plateau in the time -
dependent correlation function and has been predicted first in the
ref.\cite{Miglior}. Motivated by our dynamic studies we come here back to the
static replica approach and reformulate the theory in such a way, that we are
able to treat the freezing of the chain on internal scales, which correspond
to (Rouse) modes.

The idea is then to use the indication of the center of mass freezing and
propose on these grounds a static theory for {\it pinned} chains in random
medium.  Clearly then, the theory has to be formulated in terms of the replica
theory in such a way that the behavior {\it internal} degrees of freedom can
be treated within the same framework. Obviously more than a formulation which
is only sensitive to the overall size of the polymer must then be employed.  A
more refined theory needs then more variables than the overall size of the
chain, since the behavior of individual segments of the chain needs to be
studied in view of their freezing behavior.  Nevertheless we can expect that
within the static approach the freezing and nonergodic regimes show up through
the replica symmetry breaking (RSB) scheme \cite{Mezard,Dots}.

We reformulate in this paper the replica freezing theory and discuss the
different regimes for a Gaussian chain by making use the replica approach
embedded in the variational formalism \cite{Sasai,Takada1,Takada2,Gold1}.
Indeed we are going to show that the corresponding free energy can be
explicitly calculated in the framework of so - called one-step replica
symmetry breaking (1-RSB) scenario \cite{Mezard,Dots}. Moreover it will be
formulated more generally, that annealed and the replica symmetric theory can
be treated as special cases in a more general framework. Within this scenario
all replicas are grouped into clusters so that the resulting pattern has only
two levels: the intra-cluster overlap has a finite value whereas the
inter-cluster overlap is equal to zero.  The first special case, when the
cluster contains only one self-overlapping replica, corresponds to the replica
symmetrical (RS) scenario and in principle leads back to results obtained in
ref.\cite{Edwards}. We will show that for such situations a critical value for
the disorder exists $\Delta_c \sim N^{-2+d/2}$,which corresponds to the Harris
criterion (see \cite{Miglior} and references therein). For stronger disorder
the chain is captured by the disorder potential and the chain gyration radius
(resulting from the interplay between the entropical and energetical terms) is
scaled as $R_{\rm g} \sim (1/\Delta)^{1/(4 - d)}$. This is basically a
background for the dynamic localization of the center of mass of the chain and
the chain collapse to its localization radius $R_{\rm g}$.

In the general case of 1-RSB scenario the free energy is a functional of the
mean - square beads (inter-replica) deviation within a frozen state
$\overline{{\cal D}^2}$. This corresponds to an order parameter which
characterizes the freezing of the internal degrees of freedom inside the chain. We will show that the solution for
$\overline{{\cal D}^2}$ arises through the 1-st order phase transition (first
as a metastable state) at some critical disorder $\Delta_A$ which is similar
to critical temperature $T_A$ discussed in ref. \cite{Sasai,Takada1,Takada2}.
It is of interest that for reasonably long chains $\Delta_c < \Delta_A$, so
that the center of mass is localized first and then with increasing $\Delta$
internal degrees of freedom become frozen.

\section{Replicas and variational approach}
\subsection{Model}

We consider a Gaussian chain which is characterized by the $d$-dimensional
vector - function ${\bf R}(s)$, where $1 \le s \le N$ and labels beads of
chain. The chain is embedded in a quenched random potential ,$V\{{\bf
  R}(s)\}$, so that the whole Hamiltonian take the form
\begin{eqnarray}
  H= A \sum_{s=1}^{N}\left[\nabla_{s}{\bf
     R}(s)\right]^{2} + \sum_{s=1}^{N} V\{{\bf
   R}(s)\} 
 \quad,
 \label{Hamilton}
\end{eqnarray}
where $A = d/2b^2$ , $b$ is the Kuhn segment length, $N$ is the length 
of chain and the finite difference $\nabla_{s}{\bf R}(s) = {\bf R}(s +
1) - {\bf R}(s)$. We will use below the units of measurement where $k_B 
T=1$. The quenched random potential $V\{{\bf R}(s)\}$ is assumed to be 
Gaussian distributed with a short-ranged correlator
\begin{eqnarray}
 \left < V({\bf r})V({\bf r'}) \right>  &=& \frac{\Delta}{(2\pi
   a^2)^{d/2}} 
   \exp\left[- \frac{({\bf r} - {\bf r'})^2}{2 a^2} \right]\nonumber\\  
 &\to& \Delta \delta^{(d)}( {\bf r} - {\bf r'}), 
 \label{Delta}
\end{eqnarray}
where the dispersion $\Delta$ is the main control parameter of the
problem. The correlation length $a$ in eq.(\ref{Delta}) plays the role
of the spatial resolution or a minimal scale length. The second line
in eq.(\ref{Delta}) corresponds to the limit $a \to 0$. 

\subsection{Replicated variational method}

As pointed out in earlier for the quenched disorder problem the
free energy must be averaged over the random field $V\{{\bf R}(s)\}$,
i.e. $\left[F\right]_{\rm av} = - \left[\ln Z \right]_{\rm av}$, where 
$\left[\dots \right]_{\rm av}$ stands for the average over the
quenched field and $Z$ is the partition function. 

In order to average $\ln Z$ over $V$ we use the replica trick
\cite{Mezard,Dots}. After introducing $n$ copies of the chain and
averaging over $V$ with the correlator (\ref{Delta}) we obtain
\begin{eqnarray}
\left[F\right]_{\rm av} =  - \lim_{n
  \to 0} \frac{\ln\left[Z^n\right]_{\rm av}}{n} = \lim_{n
  \to 0}\frac{F_{\rm eff}}{n} \quad,
 \label{replica}
\end{eqnarray}
where the replicated partition function reads
\begin{eqnarray}
\left[ Z^n \right]_{\rm av} = \int \prod_{a=1}^{n} D {\bf
  R}(s)\exp\biggl\{ - A \sum_{a=1}^{n}\sum_{s=1}^{N}
\left[\nabla_{s}{\bf R}_a(s)\right]^2 +
\frac{\Delta}{2}\sum_{a=1}^{n}\sum_{b=1}^{n}\sum_{s=1}^{N}\sum_{s'=1}^{N}\delta\left({\bf
    R}_a(s) - {\bf R}_b(s')\right)\biggr\}\quad,
\label{partition}
\end{eqnarray}
and  ${\bf R}_a(s)$ with $a=1,2,\dots n$ is the replicated vector -
function. The effective free energy 
\begin{eqnarray}
F_{\rm eff} = - \ln\left[Z^n\right]_{\rm av} = - \ln \int \prod_{a=1}^{n} D{\bf R}_a(s)\exp\left\{- H_{\rm eff}\right\}
\label{F_eff}
\end{eqnarray}
with  the effective Hamiltonian having the form
\begin{eqnarray}
 H_{\rm eff} = A \sum_{a=1}^{n}\sum_{s=1}^{N}\left[\nabla_{s}{\bf
     R}_a(s)\right]^2 - \frac{\Delta}{2}\sum_{a,b=1}^{n}\sum_{s,s'=1}^{N}\delta\left({\bf
    R}_a(s) - {\bf R}_b(s')\right)\quad.
\label{H_eff}
\end{eqnarray}
Because eq.(\ref{partition}) is not amenable for the exact calculation 
we use the replicated variational method
\cite{Sasai,Takada1,Takada2}. This method is based on the inequality
\begin{eqnarray}
F_{\rm var} \equiv F_{\rm tr} + \left< H_{\rm eff} - H_{\rm tr}\right> 
\ge F_{\rm eff}\quad,
\label{F}
\end{eqnarray}
We take the trial Hamiltonian $H_{\rm tr}$ in the following bilinear
form
\begin{eqnarray}
 H_{\rm tr} = A\sum_{a=1}^{n}\sum_{s=1}^{N}\left[\nabla_{s}{\bf
     R}_a(s)\right]^2  + B \sum_{a=1}^{n}\sum_{s=1}^{N} {\bf
     R}_a^2(s) + C \sum_{s=1}^{N}\sum_{a \ne b}^{n} q_{ab} \left[{\bf
       R}_a(s) - {\bf R}_b(s)\right]^2
\label{H_tr}
\end{eqnarray}
with the trial free energy
\begin{eqnarray}
 F_{\rm tr} = - \ln\int \prod_{a=1}^{n} D{\bf R}_a(s)\exp\left\{- H_{\rm tr}\right\}
\label{F_tr}
\end{eqnarray}
 It is worthwile to  note that because of the presence of the
  second term in
  eq.(\ref{H_tr}) the ``trial'' Hamiltonian does not preserve the
  translational symmetry, ${\bf R}_a(s) \to {\bf R}_a(s) + {\bf C}$,
  which is respected by the effective Hamiltonian
  (\ref{H_eff}). Nevertheless it does not damage our results because
  the chain ultimately is captured by the disorder potential at
  $\Delta > \Delta_c$ in some part of the space (see Sec. III
  A). Since this part of the space is not specified one should average 
  over the chain locations or (which is equivalent) over the random
  field realizations (see e.g. Sec. 2.1 in ref.\cite{Lifshits}). 
We  might expect that the optimization of $F_{\rm var}$ with respect to $B$
and $C$ leads to a good approximation for $F_{\rm eff}$, i.e. the optimized
$F_{\rm var}^{*} \approx F_{\rm eff}$. After that with the use of
eq.(\ref{replica}) we can estimate the free energy $\left[F\right]_{\rm av} =
\lim_{n \to 0}F_{\rm var}^{*}/n$.  In eq.(\ref{H_tr}) $B$ is a variational
parameter conjugated to the chain mean-square gyration radius
$\overline{R_{\rm g}^2}$ which is responsible for the collapse transition. The
second variational parameter, $C$, is conjugated to the mean - square
inter-replica deviations $\overline{{\cal D}^2}$ which characterizes the
freezing transition. The form of the $q_{ab}$ - matrix is related with the
pattern of 1-RSB scenario which we have assumed here. This pattern looks as
follows: all $n$ replicas are divided into $n/m$ clusters each of which has
the size $m$, so that $q_{ab}$ is 1 if $a$ and $b$ $ (a \ne b)$ belong to the
same cluster and 0 otherwise. The averaging $<\dots>$ in eq.(\ref{F}) means
the expectation value with the trial Hamiltonian $H_{\rm tr}$.

\subsection{Free energy calculation}

The variational free energy $F_{\rm var}$ can be represented in the
form
\begin{eqnarray} 
F_{\rm var} = F_{0}  +  F_{1} \quad,
\label{01}
\end{eqnarray}
where the entropy term $F_{0}$ is
\begin{eqnarray} 
F_{0} = F_{\rm tr}  - \left< H_{\rm tr} \right>  +  A \left<\sum_{a=1}^{n}\sum_{s=1}^{N}\left[\nabla_{s}{\bf
     R}_a(s)\right]^2\right>
\label{entropy}
\end{eqnarray}
and the interaction part $F_1$ reads
\begin{eqnarray} 
F_1 = - \frac{\Delta}{2}\sum_{a,b=1}^{n}\sum_{s,s'=1}^{N}\Bigl<\delta\left({\bf
    R}_a(s) - {\bf R}_b(s')\right)\Bigr>\quad.
\label{interaction}
\end{eqnarray}

Let us calculate first $F_{\rm tr}$. For this purpose we must first
diagonalize the quadratic form (\ref{H_tr}) with respect to the
replica index. The coefficient matrix of the last part of the
quadratic form (\ref{H_tr}) has the $m \times m$ - blocks over
diagonal. It can be seen easily  that this matrix has $(n/m)$ - modes
with the eigenvalue 0 and $(n - n/m)$ modes with the eigenvalue $2m$
\cite{Sasai,Takada1,Takada2}. Then the eq.(\ref{F_tr}) take the form
of Gaussian integral
\begin{eqnarray}
F_{\rm tr} = - \ln \int \prod_{a=1}^{n} D{\bf R}_a(s)\exp\left\{ - \sum_{a=1}^{n}\sum_{s=1}^{N}\left[ A
  \left[\nabla_{s}{\bf R}_a(s)\right]^2 +
  B  {\bf R}_a^2(s) + C
   \Lambda_a {\bf R}_a^2(s)\right]\right\}\quad,
\label{Lambda}
\end{eqnarray}
where $\Lambda_a = 0$ for $1 \le a \le (n/m)$ and $\Lambda_a = 2m$ for 
$(n/m) + 1  \le a \le n$. By making use the  results of calculation  for the harmonic
oscillator functional integral \cite{Klein} we have
\begin{eqnarray}
F_{\rm tr} = - \frac{d}{2}\left(\frac{n}{m}\right)
\ln\left[\frac{\sinh(\lambda_{+})}{\sinh(N \lambda_{+})}\right] - 
\frac{d}{2}\left(n - \frac{n}{m}\right)
\ln\left[\frac{\sinh(\lambda_{-})}{\sinh(N \lambda_{-})}\right]
+ \frac{(N - 1)d n}{2} \ln A \quad,
\label{F_tr1}
\end{eqnarray}
where $\lambda_{+}$  and $\lambda_{-}$ can be found from relations
\begin{eqnarray} 
\frac{1}{2}\sqrt{\frac{B}{A}}  &=&
\sinh\left(\frac{\lambda_{+}}{2}\right)\nonumber\\
\frac{1}{2}\sqrt{\frac{B + 2mC}{A}}  &=&
\sinh\left(\frac{\lambda_{-}}{2}\right) \quad.
\label{eigenvalue}
\end{eqnarray}
Calculation of $\left< H_{\rm tr}\right>$ can be made by using the
scaling arguments \cite{Takada1}. Because ${\bf R}_a(s)$ are dummy
variables $F_{\rm tr}$ is not changed if we scale ${\bf R}_a(s) \to
\sqrt{z} {\bf R}_a'(s)$ and $\left< H_{\rm tr}\right> =
\left[\left(\partial/\partial z\right) F_{\rm tr}\right]_{z = 1}$. But 
according to eq.(\ref{eigenvalue})  $\lambda_{+}$  and $\lambda_{-}$  does
not depend from the rescaling of coefficients $A, B$ and $C$. The only 
part of $F_{\rm tr}(z)$ which depends from $z$ has the form $(N - 1) d 
n \ln (z A)/2$. After that we have
\begin{eqnarray}
\left< H_{\rm tr}\right> = n \frac{d}{2} (N - 1)\quad.
\end{eqnarray}
For the calculation of 
\begin{eqnarray}
A\left<\sum_{a=1}^{n}\sum_{s=1}^{N}\left[\nabla_{s}{\bf
      R}_a(s)\right]^2\right> = A \frac{\partial}{\partial A} F_{\rm tr} 
\end{eqnarray}
we make use eq.(\ref{F_tr1}). The results reads
\begin{eqnarray}
A\left<\sum_{a=1}^{n}\sum_{s=1}^{N}\left[\nabla_{s}{\bf
      R}_a(s)\right]^2\right> = n \frac{d}{2} (N - 1) &-&
\left(\frac{n}{m}\right)\frac{d}{2} A \frac{\partial
  \lambda_{+}}{\partial A} \left(\coth[\lambda_{+}] - N \coth[N
  \lambda_{+}]\right)\nonumber\\
&-&\left(n - \frac{n}{m}\right)\frac{d}{2} A \frac{\partial
  \lambda_{-}}{\partial A} \left(\coth[\lambda_{-}] - N \coth[N
  \lambda_{-}]\right)\quad.
\label{nabla}
\end{eqnarray}
For the very long chain, $N \gg 1/\lambda{\pm} ,\quad \sinh (N \lambda_{\pm}) \approx
(1/2) \exp(N \lambda_{\pm}), \coth(N \lambda_{\pm}) \approx 1$ and
eqs. (\ref{F_tr1}) - (\ref{nabla}) can be substantially simplified. For 
  example the trial free energy (\ref{F_tr1}) takes the form
\begin{eqnarray}
F_{\rm tr} = \left(\frac{n}{m}\right)\frac{d N}{2} \lambda_{+} +
\left(n - \frac{n}{m}\right)\frac{d N}{2} \lambda_{-} \quad.
\label{F_simple}
\end{eqnarray}
After straightforward calculation the entropy term (\ref{entropy})
becomes
\begin{eqnarray}
F_{0} = \left(\frac{n}{m}\right)\frac{d N}{2}\left[ \lambda_{+} -
  \frac{1}{\sqrt{1 + \frac{4 A}{B}}}\right] + n\left(1 - \frac{1}{m}\right)\frac{d N}{2} \left[ \lambda_{-} -
  \frac{1}{\sqrt{1 + \frac{4 A}{B + 2mC}}}\right]\quad.
\label{entropy1}
\end{eqnarray}
The calculation of the interaction term $F_1$ (see eq.(\ref{interaction})) is
more complicated because it involves the replicated Green function of
the bilinear form (\ref{H_tr}). Employing the exponential
representation of the $\delta$ - function in (\ref{interaction}) we have
\begin{eqnarray}
F_1 = - \frac{\Delta}{2}\sum_{a,b=1}^{n}\sum_{s,s'=1}^{N}\int\frac{d^d 
  k}{(2\pi)^d} \:\: Q_{ab} (s, s'; {\bf k}) \quad,
\label{interaction1}
\end{eqnarray}
where the replicated correlator
\begin{eqnarray}
 Q_{ab} (s, s'; {\bf k}) &=& \Bigl< \exp\left\{i {\bf k}\left[{\bf
       R}_a(s) - {\bf R}_b (s')\right]\right\}\Bigr>\nonumber\\
&=&\exp\left\{ - \frac{k^2}{4A}\left[ G_{aa}(s,s) + G_{bb}(s', s')  - 2
    G_{ab}(s, s')\right]\right\}\quad.
\label{correlator_Q}
\end{eqnarray}
In eq.(\ref{correlator_Q}) the Green function 
\begin{eqnarray}
 G_{ab}(s, s') = \left[{\cal H}^{-1}\right]_{ab} (s, s')
\label{Green_function}
\end{eqnarray}
is the result of  inversion of the coefficient matrix in the quadratic 
form
\begin{eqnarray}
H_{\rm tr} = \sum_{c,d=1}^{n}\sum_{s,s'=1}^{N}{\bf R}_c (s){\cal
  H}_{cd}(s, s') {\bf R}_d (s') \quad.
\label{quadratic}
\end{eqnarray}
The explicit calculation of $Q_{ab}$ and $F_1$ is given in the
Appendix A. It turns out, that the $Q_{ab}$ take different forms in the
respective elements of the matrix. 
The diagonal elements $Q_{a a}$ read
\begin{eqnarray}
Q_{a a}(s, s'; {\bf k}) &=& \exp\left\{ - \frac{k^2}{4A}\left[\overline{g}(s,
  s) + \overline{g}(s', s') - 2 \overline{g}(s,
  s')\right]\right\}\nonumber\\
&=&\exp\left\{ - \frac{k^2}{4A m \sinh (\lambda_{+})} \left[ 1 - e^{-
      \lambda_{+}|s - s'|}\right]  - \frac{k^2(m-1)}{4A m \sinh (\lambda_{-})} \left[ 1 - e^{-
      \lambda_{-}|s - s'|}\right] \right\} \quad.
\label{Q_1}
\end{eqnarray}
 The off - diagonal elements of $Q_{a b}$ inside the blocks are calculated to be
\begin{eqnarray}
Q^{<}_{a b}(s, s'; {\bf k}) &=& \exp\left\{ -
  \frac{k^2}{4A}\left[\overline{g}(s,s) + \overline{g}(s', s') - 2
    g(s, s')\right]\right\}\nonumber\\
&=&\exp\left\{ - \frac{k^2}{4A m \sinh (\lambda_{+})} \left[ 1 - e^{-
      \lambda_{+}|s - s'|}\right]  - \frac{k^2}{4A m \sinh
    (\lambda_{-})} \left[ m - 1 + e^{-
      \lambda_{-}|s - s'|}\right] \right\} \quad.
\label{Q_2}
\end{eqnarray}
Finally the off - diagonal elements of  $Q_{a b}$ outside the blocks can
be written as 
\begin{eqnarray}
Q^{>}_{a b}(s, s'; {\bf k}) &=& \exp\left\{ -
  \frac{k^2}{4A}\left[\overline{g}(s,s) + \overline{g}(s',
    s')\right]\right\}\nonumber\\
&=&\exp\left\{ - \frac{k^2}{4A}\left[ \frac{1}{ m \sinh (\lambda_{+})} 
    + \frac{m - 1}{ m \sinh (\lambda_{-})}\right]\right\} \quad.
\label{Q_3}
\end{eqnarray}
In eqs. (\ref{Q_1}) - (\ref{Q_3}) $\overline{g}$ and $g$
stand respectively for the diagonal and off - diagonal elements of the 
Green - function.
In addition we have to see how many elements $Q_{a b}$  of each type exist for
the present problem.  It is easily seen
that the number of diagonal elements is $n$. The number of elements
$Q^{<}_{a b}$ is $n(m - 1)$ and similarly the number of elements of the 
$Q^{>}_{a b}$ is $n(n - m)$.

The final form of $F_{1}$ is given by 
\begin{eqnarray}
F_1 (n, m) &=&  \frac{\Delta}{2 \pi^{d/2} b^d} \sum_{s, s'=1}^{N}
\biggl\{ n \left[\frac{1 - e^{-\lambda_{+}|s - s'|}}{m \sinh
    (\lambda_{+})} + \frac{m - 1}{m} \:\: \frac{ 1 - e^{- \lambda_{-}|s -
      s'|}}{\sinh(\lambda_{-})}\right]^{- d/2}\nonumber\\
&+& n(m - 1)\left[\frac{1 - e^{-\lambda_{+}|s - s'|}}{m \sinh(\lambda_{+})} + 
\frac{m -  1 + e^{- \lambda_{-}|s -
    s'|}}{m \sinh(\lambda_{-})}\right]^{- d/2}\nonumber\\
&+& n(n - m)  \left[\frac{1}{m \sinh(\lambda_{+})} + \frac{m - 1}{m}
  \:\: \frac{1}{\sinh(\lambda_{-})}\right]^{- d/2} \biggr\}
\label{interaction_final}
\end{eqnarray}
and the complete variational free energy is then given by the sum of all the
individual contributions.

\subsection{Order parameters}

As have been mentioned in the Introduction there are at least two
types of phenomena which we face while the disorder strength $\Delta$
is growing. At some critical $\Delta_c$ the chain is captured by the
disorder potential, so that the chain center of mass becomes localized 
and the mean - squared gyration radius $\overline{R_{\rm g}^2}$ is defined
only by disorder. In this case $\overline{R_{\rm g}^2}$  play the
role of the order parameter and can be expressed in terms of
variational parameters $B, C$ as follow
\begin{eqnarray}
\overline{R_{\rm g}^2} &=& \frac{1}{n N} \left<
  \sum_{c=1}^{n}\sum_{s,s'=1}^{N}{\bf R}_c^2(s)\right> =  \frac{1}{n
  N} \frac{\partial}{\partial B} \: F_{\rm tr}\nonumber\\
&=& \frac{d}{2}\left\{\frac{1}{m}\: \frac{1}{B \sqrt{1 + \frac{4 A}{B}}} 
 +  \frac{m - 1}{m} \frac{1}{(B + 2mC) \sqrt{1 + \frac{4 A}{B + 2mC}}}
\right\} \quad,
\label{gyration_R}
\end{eqnarray}
where we have used eqs.(\ref{eigenvalue}) and (\ref{F_simple}).
In the Sec.III we will show that this quenched problem can be treated
within RS - scenario, i.e. at $m = 1$.

On the other hand within 1-RSB scenario the (meta)stable solution for
the variational parameter $C$ appears at some critical
$\Delta_A$. Parameter $C$ is conjugated to the mean - square beads
deviation within a frozen state $\overline{{\cal D}^2}$ which can be
written as 
\begin{eqnarray}
\overline{{\cal D}^2} &=& \frac{1}{n(m - 1)N} \sum_{a,b=1}^{n}\sum_{s =
  1}^{N} \left< \left[{\bf R}_a(s) - {\bf R}_b(s)\right]^2\right> =
\frac{1}{n(m - 1)N} \: \frac{\partial}{\partial C} \: F_{\rm
  tr}\nonumber\\
&=& \frac{d}{(B + 2mC) \sqrt{1 + \frac{4A}{B + 2m C}}} \quad.
\label{deviation_D}
\end{eqnarray}
The order parameter  $\overline{{\cal D}^2}$ is an inter-replica value
which characterizes the mean - squared  beads localization length and
freezing. On the other hand, $\overline{{\cal D}^2}$ can be also
interpreted as the long - time limit correlation, i.e. $\overline{{\cal D}^2} = \lim_{t \to
  \infty} N^{-1} \sum_{s = 1}^{N}\left< \left[{\bf R}(s,t) - {\bf
      R}(s,0)\right]^2\right>$. We will
consider this in Sec. IV.

\section{Center of mass localization and collapse}

\subsection{RS - scenario}

We will now investigate the quenched problem when the chain is captured by 
the disorder potential and collapse, so that the chain size is
determinated only by the disorder strength $\Delta$. The schematic
pictorial representation of this situation is given in Fig.1.

\begin{figure}[ht]
 \begin{center}
\setlength{\unitlength}{\linewidth}  
 \begin{picture}(1.0,0.45)(0,0) 
  \put(0,0){\includegraphics[width=0.39\linewidth]{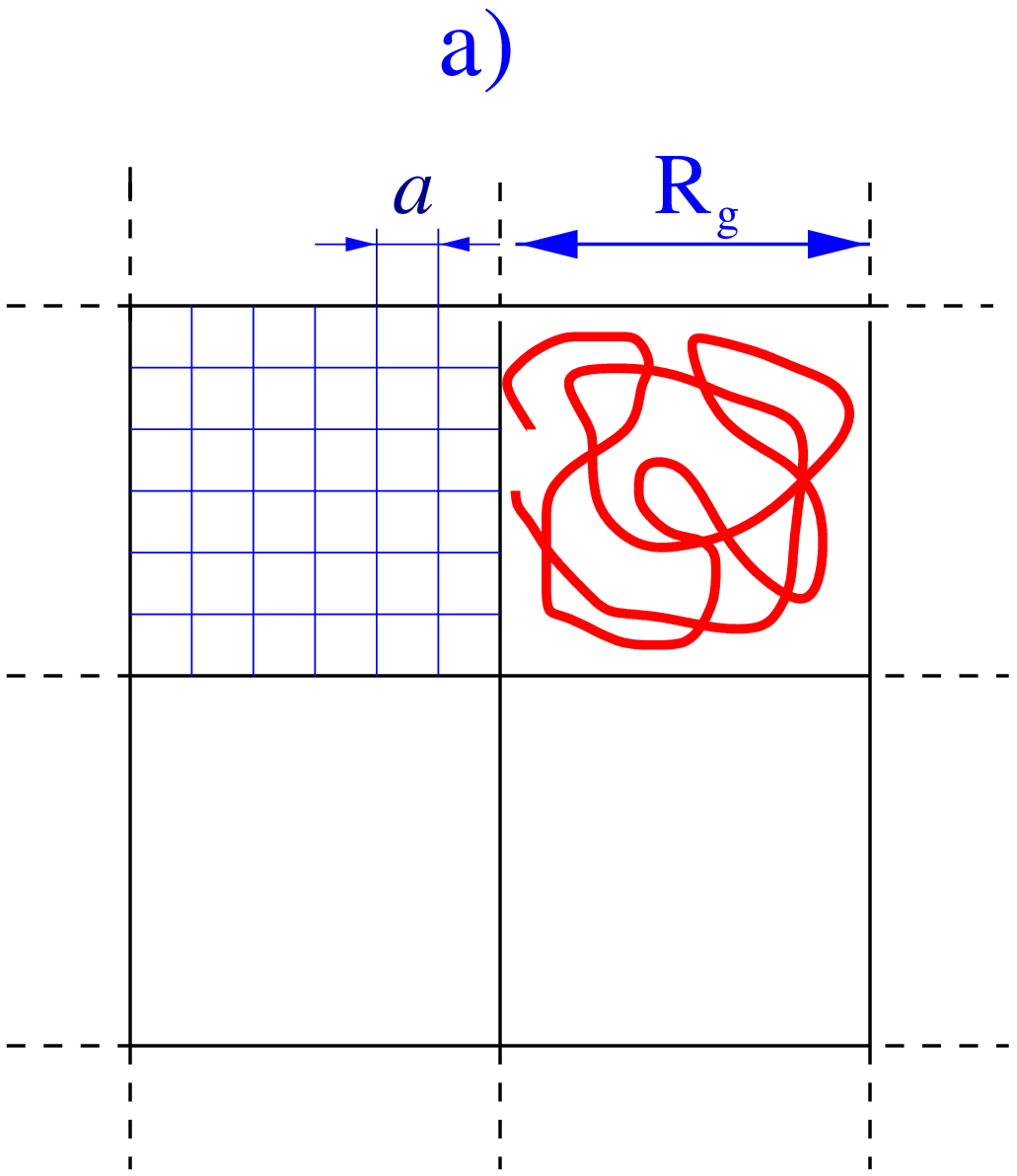}}
 \put(0.6,0.05){\includegraphics[width=0.26\linewidth]{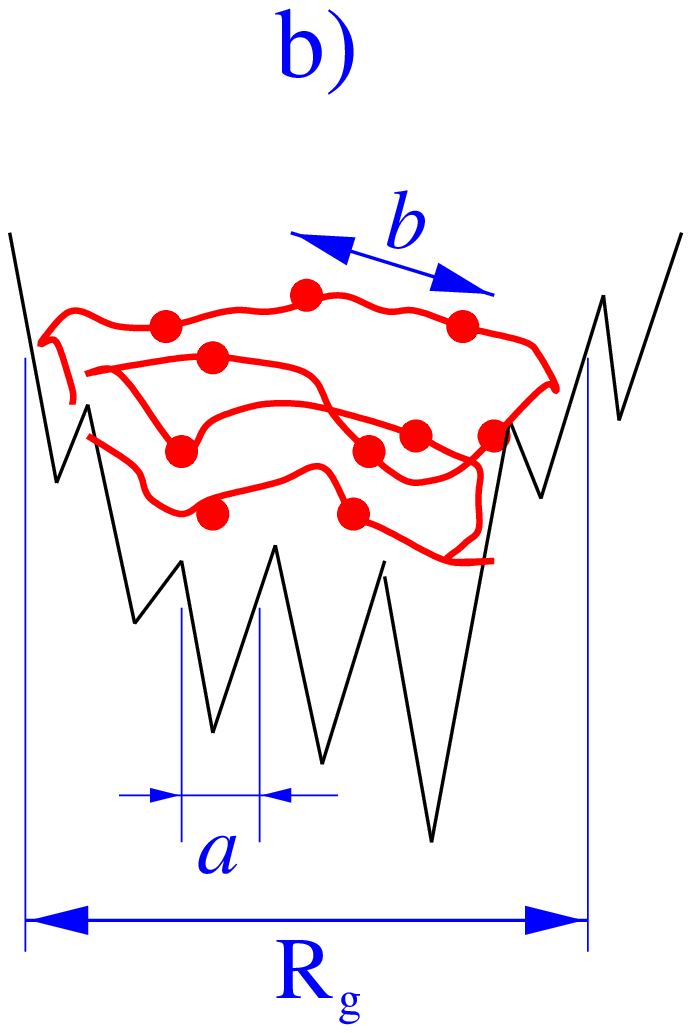}}
    \end{picture} 
\caption{\label{Cartoon} a) At the critical disorder $\Delta_c$ chain
  is captured by the random potential in some part of the space,
  i.e. becomes localized. b) The characteristic scales of the problem: 
  $a$ is the correlation length of the random potential (or the
  spatial resolution), $b$ is the Kuhn segment length, $R_{\rm g}$
  is the chain gyration radius. Scales  satisfy the
  following relations: $a \ll b \ll R_{\rm g}$ }
  \end{center} 
\end{figure}

For the purpose of calculations let us take the RS - limit (i.e. $n
\to 0$ and $m = 1$) in the interaction (\ref{interaction_final}) and
entropy (\ref{entropy1}) terms.
Using  eq.(\ref{interaction_final}) and
switching from the summation to the integration over the contour
variable $s$ leads us (after using eq.(\ref{replica})) to the following 
RS - interaction free energy part
\begin{eqnarray}
F_{\rm RS}^{(1)} = - \frac{\Delta}{2\pi^{d/2}} N
\int\limits_0^{\infty} ds
\left\{\frac{1}{\left[\overline{R^2}(s)\right]^{d/2}}\quad  - \quad
  \frac{1}{\left[\overline{R^2}(\infty)\right]^{d/2}}\right\} \quad,
\label{RS}
\end{eqnarray}
where the mean - square distance
\begin{eqnarray}
\overline{R^2}(s) = \frac{b^2}{\sinh (\lambda_{+})} \left(1 -
  e^{-\lambda_{+} s}\right) \; .
\label{mean_square}
\end{eqnarray}
Note that this form of the polymer radius corresponds to a simple solution of
a Gaussian chain in a  harmonic potential. The chain statistics is
still Gaussian on the interval $0 < s < 1/\lambda_{+}$ while the
collapsed configurations can be revealed on the larger contour length
$1/\lambda_{+} < s < N$. As a result (and because of $N \lambda_{+}
\gg 1$) the chain size is completely independent of its total contour
length $N$.
Note that the eigenvalue $\lambda_{+}$ in the limit $B \ll A$ (see below) is
small and reads
\begin{eqnarray}
\lambda_{+} \approx \sqrt{\frac{B}{A}} \ll 1 \quad.
\label{sqrt}
\end{eqnarray}
The free energy (\ref{RS}) is  typical for the variational method
and corresponds to the one - loop approximation. For the problem under 
discussion it was obtained first in \cite{Edwards}. Unfortunately the
integral in eq.(\ref{RS}) has the divergency at small $s$ and must be
regularized. The reason for this divergency lies in the contradiction
between the Gaussian statistics on the interval $0 < s <
1/\lambda_{+}$ and the effective segment - segment attraction which
shows up on all scales within the variational method (see
eq.(\ref{RS})). The integral regularization can be obtained by the
reasonable coarse - graining. Namely, we should  apply our variational
method to the string of statistically independent segments of length
$l \approx 1/\lambda_{+}$ (see Appendix B for more details). On this
coarse - grained level of description the variational method with only 
one parameter $B$ (as opposed to the case when $B(s)$ is a trial s -
dependent function \cite{Miglior1})  is getting self - consistent and the integral in
eq.(\ref{RS}) can be regularized. It should be mentioned another
approach \cite{Gold1} regarding similar point.

RS - entropy contribution , $F_{\rm RS}^{(0)}$, resulting from
eq.(\ref{entropy1}) in the limit $m = 1, n \to 0$ and taking into
account eqs. (\ref{eigenvalue}) and (\ref{sqrt}). The final expression for the full  RS - free
energy $F_{\rm RS}$ can be put in a following form
\begin{eqnarray}
F_{\rm RS} &=& F_{\rm RS}^{(0)} + F_{\rm RS}^{(1)}\nonumber\\
&=& \frac{d N}{4} \sqrt{\frac{B}{A}} - \frac{\Delta  N}{2 \pi^{d/2}
  b^d} \left(\frac{B}{A}\right)^{\frac{d - 2}{4}} \: I_d \quad,
\label{full_RS}
\end{eqnarray}
where we have used also eqs.(\ref{RS_1}) - (\ref{I_d}) for $F_{\rm
  RS}^{(1)}$ and $I_d$ respectively. For the mean - square gyration
radius from (\ref{gyration_R}) we get
\begin{eqnarray}
\overline{R_{\rm g}^2} \approx \frac{1}{\sqrt{A B}} \quad.
\label{gyration_R_simple}
\end{eqnarray}
Minimization of the free energy (\ref{full_RS}) with respect to $B$
leads to the solution 
\begin{eqnarray}
B^{*} \simeq A \left(\frac{\Delta}{b^d}\right)^{\frac{4}{4 - d}} \quad.
\label{B*}
\end{eqnarray}
The resulting $\overline{R_{\rm g}^2}$ becomes
\begin{eqnarray}
\overline{R_{\rm g}^2} \approx b^2 \left(\frac{b^d}{\Delta}\right)^{\frac{2}{4 - d}}
\label{gyration_in_disorder}
\end{eqnarray}
at $d < 4$. This result has been obtained first in
ref.\cite{Edwards}. We can meet the condition $B \ll A$ (see
eq.(\ref{sqrt})) if $\Delta \ll b^d$, i.e. $\overline{R_{\rm g}^2} \gg b^2$ 
(see Fig. 1b). The free energy at the minimum $B = B^{*}$ reads
\begin{eqnarray}
F_{\rm RS}\{B^{*}\} \sim - N
\left(\frac{\Delta}{b^d}\right)^{\frac{2}{4 - d}} < 0 \quad,
\label{minimum}
\end{eqnarray}
which corresponds indeed to the form the the free energy confinement $F_{conf}
\propto Nb^{2}/R_{\rm g}^{2}$ for localized chains.  As a result  we conclude 
that that the chain is captured by the disorder potential at some
critical disorder provided that $N \left(\Delta_{\rm c}/b^d\right)^{2/(4 - d)}
\simeq 1$, i.e.
\begin{eqnarray}
\Delta_c \simeq  b^d N^{-2 + \frac{d}{2}} \quad.
\label{critical}
\end{eqnarray}
This criterion has been recently obtained by us within the dynamical
consideration in ref.\cite{Miglior} in various representations. It corresponds
to the usual Harris criterion for chains in disorder. The quenched RS - free
energy (\ref{full_RS}) and the chain confinement is shown in Fig.2.
\begin{figure}[ht]
\begin{center}
\begin{minipage}{11cm}

    \centerline{\includegraphics[width=8cm,angle=270]{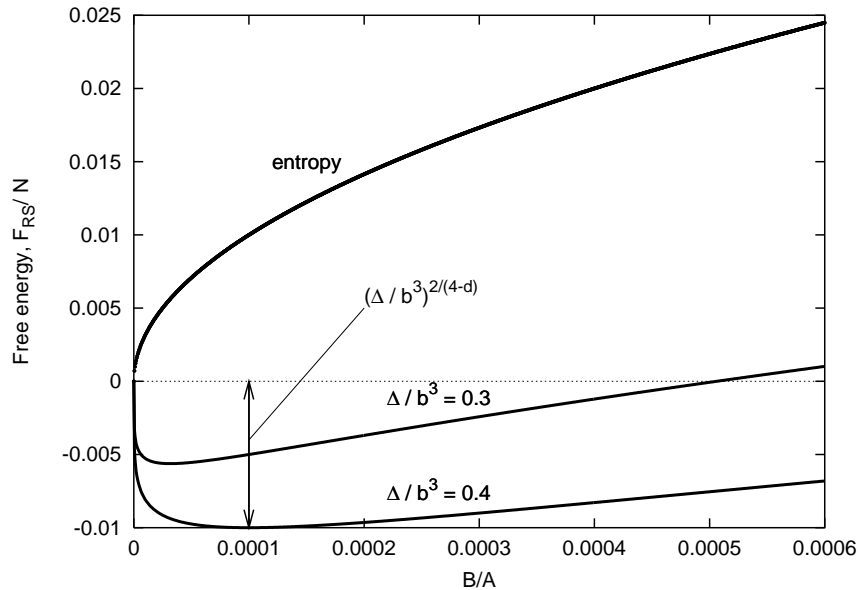}} \vspace*{10pt}

      \caption{\label{Replica_symmetrical}
         RS - free energy as a function of the variational parameter
         $B$ and different disorder strength $\Delta/b^d$. The
         confinement is guaranteed by the condition that the minimum
         depth is large enough, i.e. $N (\Delta/b^3)^{2/(4 - d)} >
         1$. The upper curve demonstrates the entropy term}
\end{minipage}
\end{center}
\end{figure}

The RS - scenario reveals two facets of the problem: the chain center of mass
localization (or confinement) which is described by the criterion
(\ref{critical}) and the collapse with the characteristic chain size given by
eq. (\ref{gyration_in_disorder}). Here it should be recorded that as
distinct from the results of ref. \cite{Cates,Natter,Gold1}  the 
expression for the chain size does  not contain the factor $\ln \Omega$ 
(where $\Omega$ is the volume of the whole random medium) since the
chain is effectively pinned for   $\Delta > \Delta_c$.

\subsection{Annealed problem}

In this case, as we have discussed in the Introduction, the chain is
fairly mobile and can span the whole system during the time of an
experiment.The chain experiences all possible quenched field
realizations and the replica trick is actually not necessary. The
corresponding interaction part of free energy results from
eq.(\ref{interaction_final}) at $n = m = 1$. The entropy term
follows from eq.(\ref{entropy1}) at  $n = m = 1$ as well. The
resulting annealed free energy reads \cite{Gold1}
\begin{eqnarray}
F_{\rm ann} = \frac{d N}{2} \sqrt{\frac{B}{A}} -
\frac{\Delta}{2\pi^{d/2} b^d} \left[\sinh(\lambda_{+})\right]^{d/2}
\sum_{s,s'=1}^{N}\: \frac{1}{\left[1 - e^{- \lambda_{+} |s -
      s'|}\right]^{d/2}} \quad.
\label{annealed}
\end{eqnarray}
By making use eqs.(\ref{mean_square}) and (\ref{sqrt}) we can
substitute the result of double summation by $N^2/R_{N}^d$ whereas the 
entropy term leads to the form $b^2 N/R_N^2$. As a result we have
\begin{eqnarray}
F_{\rm ann}  \simeq \frac{b^2 N}{R_N^2} - \Delta \frac{N^2}{R_{N}^d} 
\quad.
\label{annealed_simple}
\end{eqnarray}
This expression was obtained first in ref.\cite{Edwards,Cates}. The
minimization of eq.(\ref{annealed_simple}) leads to the following result. At
$d \ge 2$ $ F_{\rm ann} \to - \infty$ and $R_{N} \to 0$, or by cutting - off
we can state  that the chain is collapsed to a size of the order of the Kuhn
length $b$. At $d < 2$ the free energy has a minimum at $R_N \simeq
b\left(\Delta N/b^d \right)^{1/(d - 2)} \to b$ in the long chain limit if the
cut - off is imposed again. The chain crushing down is of course a result of a
Gaussian model. It is well known \cite{Miglior} that the more realistic chain
with the second $ v > 0$ and the third $w$ virial coefficients chain is
collapsed at $\Delta > v$ to the globule state with a finite density.

\section{Freezing of internal degrees  of freedom}

As shown in Sec.III the Gaussian chain is getting immobile or localized at
some critical value of disorder $\Delta_c$. This can be treated by making use
the RS - scenario. As a relevant question, we can ask what happened with
internal motions of the chain beads as the disorder $\Delta$ increases
further? It is clear that the chain still experiences the free energy
landscape and can stay  in the different minima or the conformational states.
The measure of the distance between these states is given by the order
parameter $\overline{{\cal D}^2}$ which discussed in Sec. II D.  The parameter
$\overline{{\cal D}^2}$ is used extensively for the heteropolymers, where the
landscape dominated statistics can be clearly seen \cite{Karplus}. There is a
natural mapping in the replica approach \cite{Mezard,Dots} between sets of
states and replicas. The 1-RSB - assumption about the clustering in the
replica space reflects our view of the states subdivision. The general 
pattern looks as follows. We put together in the same cluster all
states that are within a certain distance $\overline{{\cal D}^2}$ from
each other. The distance between different clusters is infinitely
large. From the expression for the entropy  (\ref{entropy1}) it can be 
seen that the fraction of states with a distance $\overline{{\cal
    D}^2}$ between them is equal to $1 - 1/m$. Fig.3 illustrates
schematically a cluster which is made up of three states: $a,b$ and $c$.
\begin{figure}[ht]
\begin{center}
\begin{minipage}{11cm}

    \centerline{\includegraphics[width=12cm]{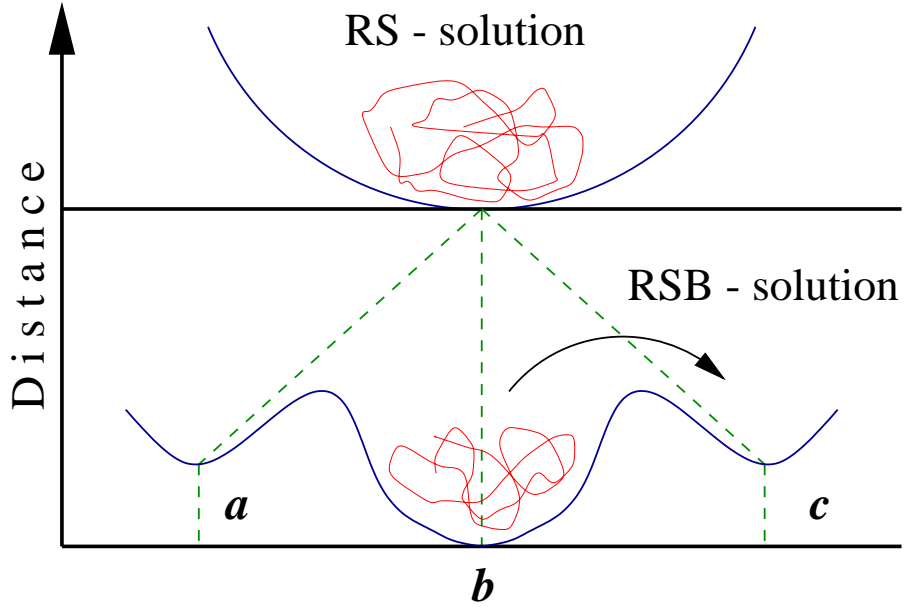}} \vspace*{10pt}

      \caption{\label{Subdivision}
         States subdivision in the space of conformations. States of
         the chain within 1-RSB - scenario (marked as $a, b, c$) form
         a cluster, so that the distance between them is equal $\overline{{\cal
    D}^2}$. To find the distance one traces back the points $a, b, c$
along the branches until they merge. The chain can move between these states.}
\end{minipage}
\end{center}
\end{figure}
In this section we will show that within 1 - RSB - scenario upon disorder
growing the parameter $\overline{{\cal D}^2}$ can take fairly small value.
This means that beads are localized in the globule interior, i.e. the system
is freezed. Indeed,  deep in the
frozen regime the change of the chain conformation can only take place on the
smallest length scales, i.e. $\overline{{\cal D}^2} \approx a^2$.

\subsection{RSB - free energy}

The frozen phase is characterized by small fluctuations around
individual minima. In this case we assume that Debye - Waller factor,
$2mC$, is large, i.e.
\begin{eqnarray}
2mC \gg A \gg B
\label{limit_gg}
\end{eqnarray}
and the order parameter given by eq. (\ref{deviation_D}) takes a
simple form
\begin{eqnarray}
\overline{{\cal D}^2} \simeq \frac{d}{2mC} \quad. 
\label{deviation_D_simple}
\end{eqnarray}
In the opposite limit of small $2mC$, i.e. at
\begin{eqnarray}
2mC \ll B \ll A
\label{limit_ll}
\end{eqnarray}
$\lambda_{-} \simeq \lambda_{+}$ and it can be seen that the free energy
functional $F_1(n,m)$ in eq.(\ref{interaction_final}) and the
entropy (\ref{entropy1}) goes back to RS - free energy as it should
be.

In the limit (\ref{limit_ll}) the order parameter $\overline{{\cal
    D}^2}$ reads
\begin{eqnarray}
\overline{{\cal D}^2} \simeq \frac{d}{B \sqrt{1 + \frac{4A}{B}}}
\simeq  \overline{R^2_g}
\label{deviation_D_small}
\end{eqnarray}
This is a natural result because in the case when the center of mass
is localized but the internal degree of freedom are still not frozen
(because of small $C$) the beads localization length is of the order
of the gyration radius. 

Let us calculate the RSB - free energy $F_{\rm RSB}$. For the sake of
simplicity we should find first of all the expressions for the entropy 
$F_{\rm RSB}^{(0)}$ (see eq.(\ref{entropy1})) and interaction $F_{\rm
  RSB}^{(1)}$ (see  eq.(\ref{interaction_final}))  terms in the both
limits, (\ref{limit_gg}) and (\ref{limit_ll}). After that we can find
a reasonable interpolating expression, which is suitable for the
theoretical investigation. These calculations are represented in
Appendix C. The overall RSB - free energy which consists of the entropy
(\ref{entropy_interpol})  and  the  interaction (\ref{RSB_interpolation})
terms is given by
\begin{eqnarray}
F_{\rm RSB}/N &=& (F_{\rm RSB}^{(0)} + F_{\rm  RSB}^{(1)}) /N\nonumber\\
&=& \frac{m - 1}{m} \: \frac{d}{2}\: \ln \left(y + 1\right) -
\frac{d  \:\Delta}{2\pi^{d/2} b^d} \:\: \frac{\lambda_{+}^{\frac{d}{2} -
    1}( m^{\frac{d}{2} + 1} - 1 ) \:\:  y}{y + \frac{m (m^{\frac{d}{2}
      + 1} - 1)}{m - 1} 2
  \lambda_{+}^{2}} \quad,
\label{RSB_final}
\end{eqnarray}
where the dimensionless Debye - Waller factor $y \equiv 2 m C b^2$.

\subsection{Freezing transition}

Now we discuss the transition related with the freezing of the
internal degree of freedom and based on the RSB - free energy
expression eq.(\ref{RSB_final}). Minimization of eq.(\ref{RSB_final})
 with respect of $y$ yields  the solution
\begin{eqnarray}
y^{*} \simeq \left(\frac{\Delta}{b^d}\right)^{\frac{6}{4 - d}} \:
\frac{m^{2}(m^{\frac{d}{2} + 1} - 1)^2}{\left(m - 1\right)^2}
\label{y*}
\end{eqnarray}
where we have used for $\lambda_{+}$ the results of RS - solution
(\ref{B*}), i.e. $\lambda_{+}\simeq (B^{*}/A)^{1/2}  \simeq
\left(\Delta/b^d\right)^{2/(4 - d)}$. When minimizing
eq. (\ref{RSB_final}) we also took into account that $y^{*}$ is close
to $y_{\rm max}$, i.e. $y^{*} \le y_{\rm max}$, where $y_{\rm max}$ is 
given by
\begin{eqnarray}
y_{\rm max} = 2 m C_{\rm max} b^2 \simeq \frac{b^2}{a^2} > 1 \quad. 
\label{y_max}
\end{eqnarray}
The stationary condition with respect to $m$, i.e. $\left(\partial F_{\rm
  RSB}/\partial m \right)_{y = y_{max}} = 0$, leads to the optimal
result for $m = m^{*}$ defined by 
\begin{eqnarray}
m^{*} \simeq  \frac{\left[\ln y_{\rm max}\right]^{\frac{2}{4 +
      d}}}{\left[\frac{\Delta}{b^d}\right]^{\frac{4}{(4 - d)(4 + d)}}} 
\quad \ge 1
\label{m*}
\end{eqnarray}
 
Solution $y = y^{*}$ appears first as a metastable one at some $\Delta = \Delta_{\rm
  A}$, which is similar to the critical temperature $T_{\rm A}$ in
heteropolymers discussed in ref.\cite{Takada1,Takada2}. This transition in a
metastable state is also of the same nature as in the $p$ - spin glass model
studied in ref.\cite{Kurchan}.

In Fig.3 we have plotted the RSB - free energy expression
eq.(\ref{RSB_final}) as a function of Debye - Waller factor $y$ at the 
different disorder strength; in doing so we have used for $m$ the optimal value given by eq.(\ref{m*}). The metastable solution shows up at
$\Delta_{\rm A} = 0.753$, so that $y^{*}$ grows with the disorder
until it hits the $y_{\rm max}$, i.e. $y^{*} \simeq y_{\rm max} =
b^2/a^2$. In our case it is  happened at $\Delta_G/b^3 \simeq 0.796$ with
$y_{\rm max} = 66.3$. At this point all beads are totally localized or 
frozen.
\begin{figure}[ht]
\begin{center}
\begin{minipage}{11cm}

    \centerline{\includegraphics[width=8cm,angle=270]{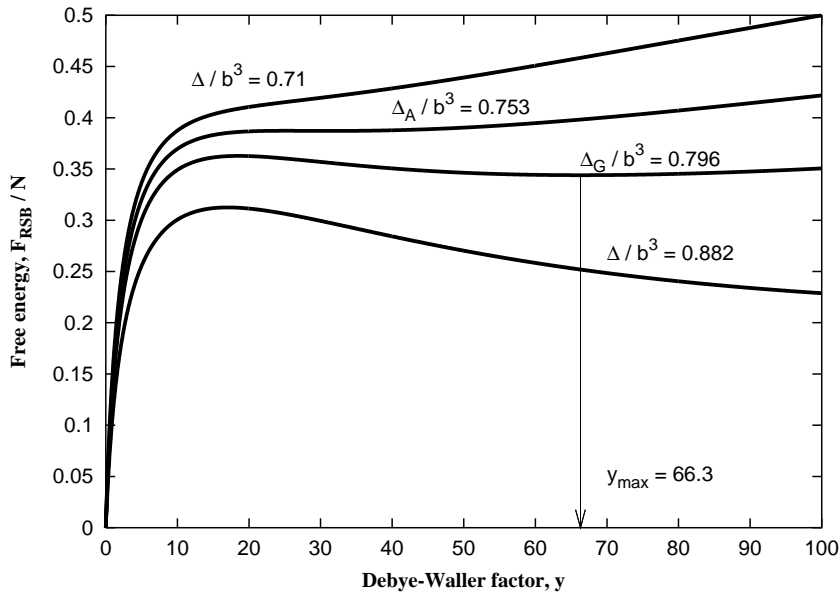}} \vspace*{10pt}

      \caption{\label{RSB}RSB - free energy at the different disorder
        $\Delta/b^3$ and fixed $y_{\rm max} = 66.3$. The arrow shows
        the minimum  position $y^{*}$ when it hits  $y_{\rm max}$,
        i.e. $y^{*} = y_{\rm max}$. The corresponding disorder
        strength $\Delta = \Delta_{\rm G} = 0.796 b^3$
         }
\end{minipage}
\end{center}
\end{figure}

It is interesting to make a link with the heteropolymer theory
\cite{Takada1,Takada2}. At $m^{*} \to 1$ the RSB - free energy $F_{\rm 
  RSB} = 0$ and eq.(\ref{m*}) immediately leads to the corresponding
expression for $\Delta = \Delta_{\rm K}$ which  meets this condition, i.e.
\begin{eqnarray}
\frac{\Delta_{\rm K}}{b^d} \simeq \left[\ln y_{\rm
    max}\right]^{\frac{4 - d}{2}} \quad.
\label{Delta_G}
\end{eqnarray}

This expression quantitatively corresponds to the 
Kauzmann temperature $T_{\rm K}$ in heteropolymers \cite{Takada1}. In
ref.\cite{Takada1} it was found that $\sigma /k_B T_{\rm K}$ (where
$\sigma$ is the variance of the random interaction) is proportional to 
the square root of the entropy loss, which is similar to
eq.(\ref{Delta_G}).  The dependence of $y^{*}$ from $\Delta$ at the
different  $y_{\rm max} = b^2/a^2$ is shown in Fig.4. The angular
points,  where $y^{*}$ hits $y_{\rm max}$, correspond to $ \Delta =
\Delta_{\rm G}$ where the chain becomes fully frozen.
\begin{figure}[ht]
\begin{center}
\begin{minipage}{11cm}

    \centerline{\includegraphics[width=8cm,angle=270]{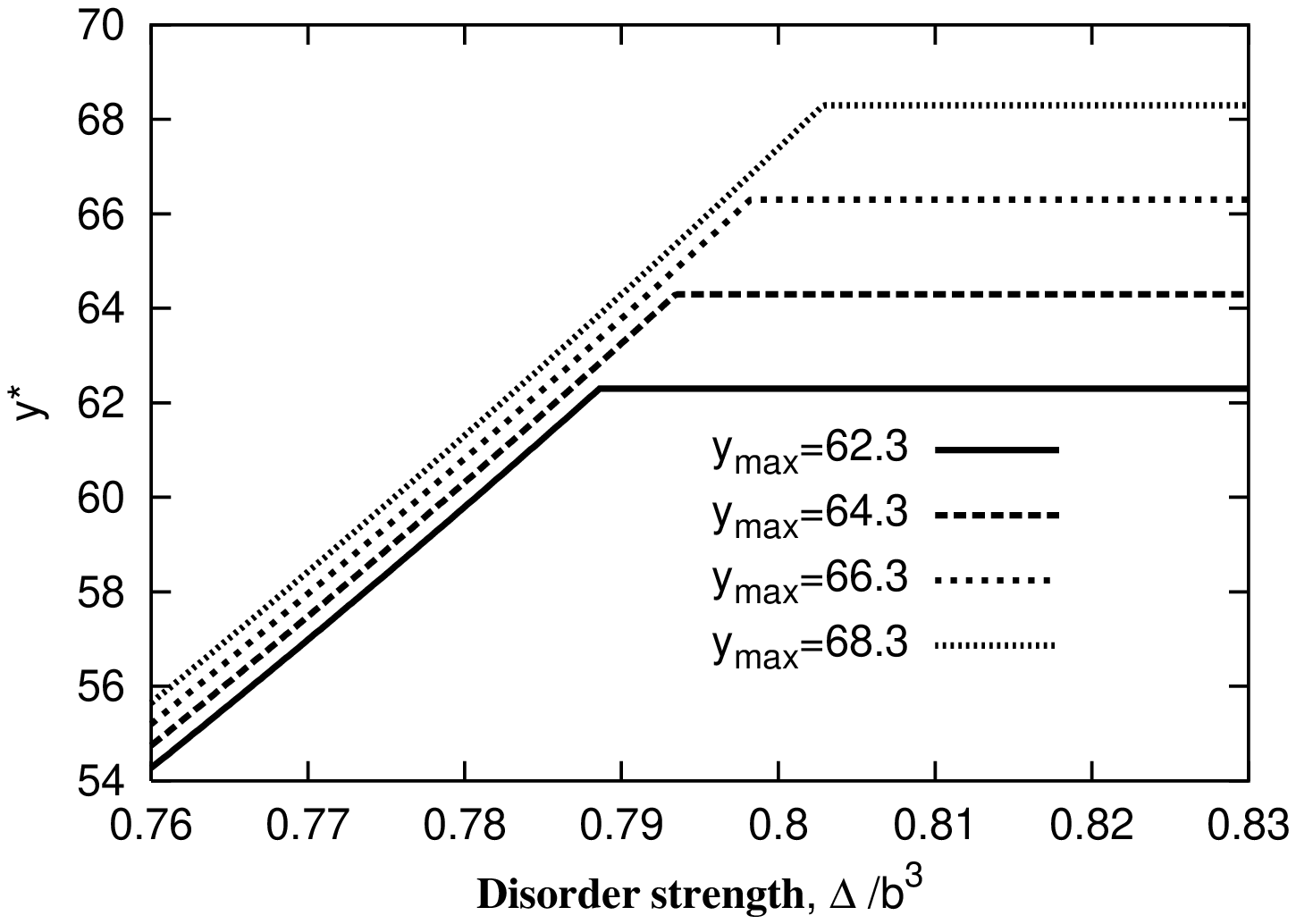}} \vspace*{10pt}

      \caption{\label{glass} Dependence of the minimum position
        $y^{*}$ from the $\Delta/b^3$ at the different $y_{\rm
          max}$. The angular point corresponds to the disorder,
        $\Delta_{\rm G}$,  when  the localization length
        $\overline{{\cal D}^2} \simeq a^2$, i.e. the chain becomes
        fully frozen. 
         }
\end{minipage}
\end{center}
\end{figure}

\begin{figure}[ht]
\begin{center}
\begin{minipage}{11cm}

    \centerline{\includegraphics[width=8cm,angle=270]{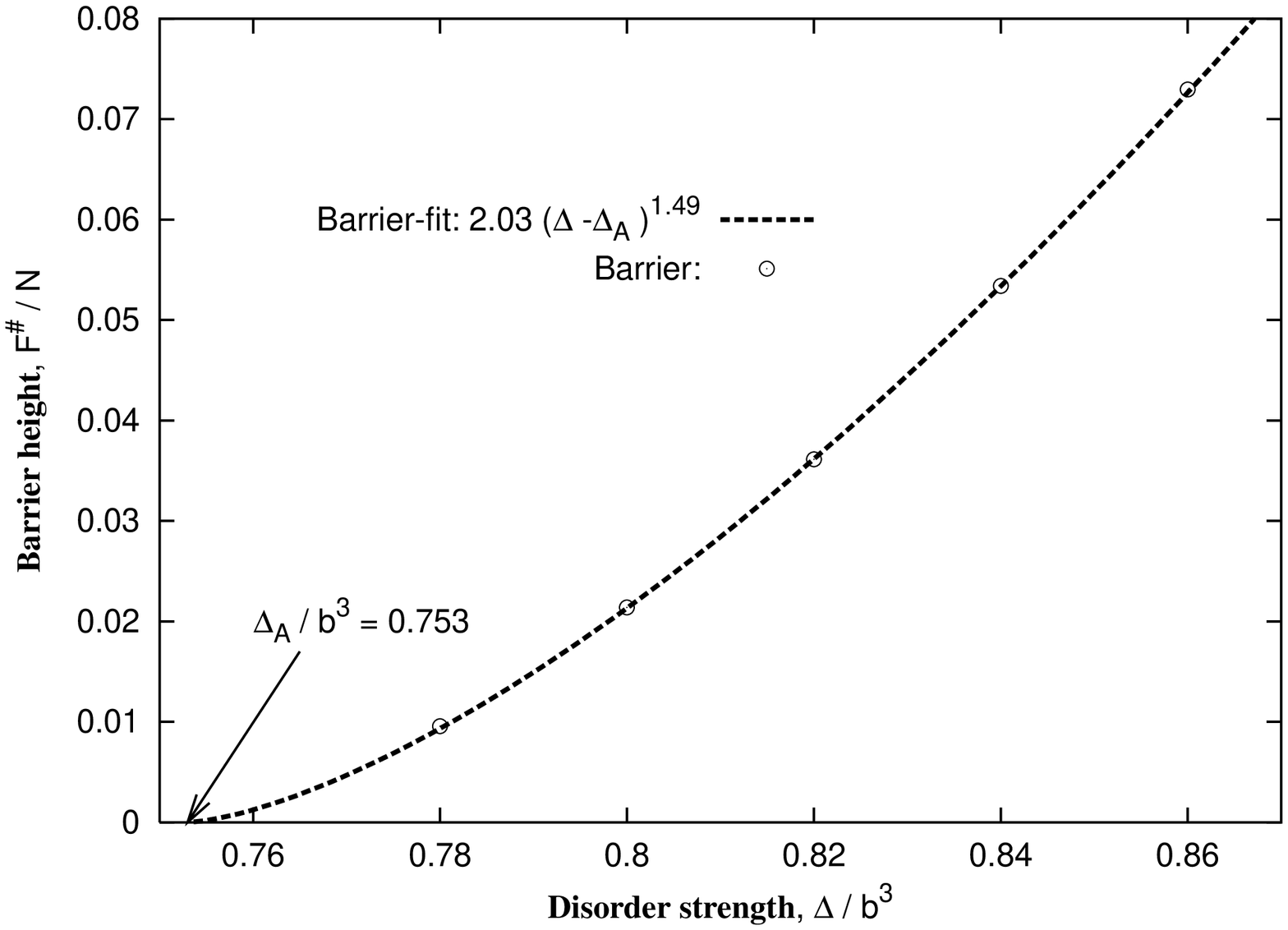}} \vspace*{10pt}

      \caption{\label{glass1} The barrier height $F^{\ddag}$ as a
        function of the disorder. The scaling form fitting corresponds 
        to the eq.(\ref{fitting}).
         }
\end{minipage}
\end{center}
\end{figure}

In the disorder interval $\Delta_{\rm A} < \Delta < \Delta_{\rm G}$
the chain is in a metastable state and the beads localization length
lies in the range $a^2 \le \overline{{\cal D}^2} \le \overline{R_{\rm g}^2}$.
The beads become localized when  the effective barrier height
between the minimum $F^{*}$ and maximum $F^{**}$ in Fig.3 is large
enough to assure the beads confinement. Fig. 5 shows the result of the 
numerical estimation of the effective barrier height $F^{\ddag}/N \equiv
\left(F^{**} - F^{*}\right)/N$. The result of the fitting can be
written as 
\begin{eqnarray}
F^{\ddag}/N  = 2.03 \left(\Delta - \Delta_{\rm A}\right)^{1.49} \quad.
\label{fitting}
\end{eqnarray}
The beads confinement condition, $F^{\ddag} > 1$, leads to the
following finite chain scaling for the transition disorder
\begin{eqnarray}
\Delta_{\rm A}(N) = \Delta_{\rm A} + \frac{0.5 b^3}{N^{0.66}}
\label{finite}
\end{eqnarray}
The eq.(\ref{finite}) means that for the reasonably long chain the
internal freezing does not depend from the chain length. The
comparison with the scaling low eq.(\ref{critical}) for the center of
mass localization critical disorder $\Delta_{\rm c}$ leads us to the
conclusion that $\Delta_{\rm c} < \Delta_{\rm A}$ at $N \gg 1$. By
this is meant that as the disorder increases the chain center of mass
is localized first at $\Delta = \Delta_{\rm c}$ and then at $\Delta =
\Delta_{\rm A}$ the internal degree of freedom start to
freeze. Furthermore at $\Delta = \Delta_{\rm G} > \Delta_{\rm A}$ the
whole system is freezed out with the beads localization length
$\overline{{\cal D}^2} \simeq a^2$.

\section{Conclusion}

We have studied the behavior of the Gaussian chain in a quenched random short
correlated field which is specified by the field dispersion $\Delta$. The
consideration is based on the variational method in replica space which was
previously used for the heteropolymer problem \cite{Sasai,Takada1,Takada2}
within the simplest 1-RSB - scenario. This assumption allows to calculate the
general free energy functional in an explicit form and consider its RS and RSB
cases.

We have shown that the RS - limit ($n \to 0$ and $m = 1$) allows to treat the
center of mass localization at some critical $\Delta_{\rm c}$ (see
eq.(\ref{critical})) as well as the chain collapse (see
eq.(\ref{gyration_in_disorder})). In line with eq.(\ref{critical}) it was
found by the methods based on the analytical Langevin dynamics calculations
\cite{Miglior} that the localization threshold for the center of mass of the
self - avoiding chain is scaled as $\Delta_{\rm c} \simeq b^d N^{-2 + \nu d}$,
where $\nu$ is the Flory exponent. More precisely, for the problem in question
there are two characteristic time scales. One of them is the time of the chain
capture $t_{\rm cap}$ by the disorder potential. Within our dynamical
calculation based on the Hartree approximation \cite{Miglior} (which in turn
is reminiscent of the mode - coupling approximation \cite{Gotz}) $t_{\rm cap}
\to \infty$. The other time scale is the equilibration time of the internal
degrees of freedom $t_{\rm eq}$. It should be realized that our results are
valid for $t_{\rm cap} \gg t_{\rm eq}$. The main conclusions of
ref.\cite{Miglior} have been recently corroborated by the direct MC -
simulations \cite{Milchev} which among other things shows a pronounced
decrease of the center of mass diffusion coefficient in the vicinity of
$\Delta_c$.  The resulting eq.(\ref{gyration_in_disorder}) corresponds to the
size of this trapped chain. Nevertheless the size itself is not sufficient to
describe the physical situation. Here we added the discussion of the internal
degrees of freedom by dividing the replicated system into $n/m$ blocks.

Then the 1-RSB - limit (i.e. at $n \to 0$ and $m = m^{*}$, where $ m^{*}$ is an
optimized value of the breaking point parameter $m$) of the free energy
expression reveals the nontrivial solution for the inter-replica value
$\overline{{\cal D}^2}$, the beads localization length. This solution appears
{\it discontinuously} at the critical disorder $\Delta = \Delta_{\rm A}$ which
corresponds to the freezing of the chain internal degree of freedom. Beads
localization length $\overline{{\cal D}^2}$ falls to $\overline{{\cal D}^2}
\simeq a^2$ with the disorder increasing in the interval $\Delta_{\rm A} <
\Delta < \Delta_{\rm G}$. These results are in a qualitative agreement with
the freezing scenario in heteropolymers \cite{Takada1,Takada2}. On the other
hand, according the findings based on the dynamical approach \cite{Miglior}
the nonergodicity function shows up (at least for the long Rouse modes) {\it
  continuously}. Moreover the critical threshold for internal Rouse modes
freezing is scaled as $\Delta_{\rm c}' \simeq b^d N^{- 0.25}$ which at first
sight can not be reconciled with eq.(\ref{finite}). At the moment it is not
clear what is the reason for this discrepancy. Presumably by going beyond the
scope of 1 - RSB - scenario to the full scale RSB -  hierarchy we could fill the
gap between dynamical and replica methods.

\section*{Acknowledgments}

We thank J. Kurchan, S. Takada and P. G. Wolynes for valuable comments.

\begin{appendix}
\section{Replicated Green function for the Gaussian trial chain}

Diagonalization of the quadratic form (\ref{H_tr}) in the replica
space leads to eq.(\ref{Lambda}). The corresponding Green function has 
two components, $G^{\pm}(s, s')$, which associated with two types of
eigenvalues. The Green function satisfies the equation of motion
\begin{eqnarray}
\left[ - \nabla_s \overline{\nabla}_s + \left(\omega^{\pm}\right)^2
\right] G^{\pm}(s, s') = \delta_{s s'} \quad,
\label{eq_motion}
\end{eqnarray}
subjected to free ends boundary conditions
\begin{eqnarray}
 \nabla_s  \left.G^{\pm}(s, s')\right|_{s = 0} =  \nabla_{s'}
 \left.G^{\pm}(s, s')\right|_{s = N} = 0 \quad.
\label{boundary}
\end{eqnarray}
In eqs. (\ref{eq_motion}) - (\ref{boundary}) the finite differences
are defined as follows: $\nabla_s G^{\pm}(s, s') \equiv  G^{\pm}(s+1,
s') -  G^{\pm}(s, s') , \overline{\nabla}_s  G^{\pm}(s, s')  \equiv
G^{\pm}(s, s') -G^{\pm}(s - 1 , s')$; the ``characteristic frequencies''
$\omega^{+} = \sqrt{B/A}$ and $\omega^{-} = \sqrt{(B + 2mC)/A}$. The
exact solution of eq. (\ref{eq_motion}) becomes the form
\cite{Sasai,Takada1}
\begin{eqnarray}
G^{\pm}(s, s') = 
\begin{cases}
\frac{\cosh \left[\lambda_{\pm}\left(s -
      \frac{1}{2}\right)\right]\cosh \left[\lambda_{\pm}\left(N -s' +
      \frac{1}{2}\right)\right] }{\sinh[\lambda_{\pm}]\sinh[N \lambda_{\pm}] },  &\text{if $s
  \le s'$}\\
\\
\frac{\cosh \left[\lambda_{\pm}\left(s' -
      \frac{1}{2}\right)\right]\cosh \left[\lambda_{\pm}\left(N - s +
      \frac{1}{2}\right)\right] }{\sinh[\lambda_{\pm}]\sinh[N \lambda_{\pm}] }, &\text{if $s
  \ge s'$} 
\end{cases}
\label{GF}
\end{eqnarray}
where $\lambda_{+}$ and $\lambda_{-}$ are defined by
eq.(\ref{eigenvalue}).

For the large chain, $N \gg 1/\lambda_{\pm}$, eq.(\ref{GF}) can be written in a much 
simpler form
\begin{eqnarray}
G^{\pm}(s, s') \approx \frac{1}{2 \sinh(\lambda_{\pm})} \exp \left\{ - 
  \lambda_{\pm} |s - s'| \right\} \quad.
\label{GF_simple}
\end{eqnarray}

In order to calculate the  Green function matrix elements $G_{a b}(s,
s')$ we recall that the $n \times n$ - matrix has the block - diagonal 
structure with each block of size $m$. Let us denote all diagonal
elements $G_{a a} = \overline{g}$, off - diagonal elements  which
belong to blocks $G_{a b} = g$ and all elements outside of blocks are
equal to zero. For the purpose of diagonalizing the matrix $G_{a b}$
we diagonalize first each $m \times m$ - block. The corresponding
characteristic equation for the eigenvalues $\chi$ is
\begin{eqnarray}
\begin{vmatrix}
\overline{g} - \chi   &   g   & \dots &   g\\
g                     &   \overline{g} - \chi  & \dots  & g\\
\vdots   &   \vdots   &  \ddots  & \vdots\\
g  &  g  &   \dots  &  \overline{g} - \chi
\end{vmatrix}
= 0
\label{characteristic}
\end{eqnarray}
It is easy to find that the eq.(\ref{characteristic}) has one solution 
\begin{eqnarray}
\chi_{+} =  \overline{g} - (1 - m) g
\label{solution1}
\end{eqnarray}
and $(m - 1)$ - solutions
\begin{eqnarray}
\chi_{-} = \overline{g} - g
\label{solution2}
\end{eqnarray}

The determinant of $G_{a b}$ is an invariant of orthogonal  transformations and
can be represented in the form
\begin{eqnarray}
\mbox{det} \left\{G_{a b}\right\} &=&
\left(\chi_{+}\right)^{\left(\frac{n}{m}\right)}\left(\chi_{-}\right)^{\left(n 
    - \frac{n}{m}\right)}\nonumber\\
&=&\left(G^{+}\right)^{\left(\frac{n}{m}\right)}\left(G^{-}\right)^{\left(n 
    - \frac{n}{m}\right)} \quad,
\label{det}
\end{eqnarray}
which leads to the identity $\chi_{+} = G^{+}$ and $\chi_{-} = G^{-}$
and allows to wright down
\begin{eqnarray}
G^{+} &=&  \overline{g} - (1 - m) g \nonumber\\
G^{-} &=& \overline{g} - g
\label{relation}
\end{eqnarray}
As a result the expressions for non - zero matrix - elements read
\begin{eqnarray}
g &=& \frac{1}{m} \left[ G^{+} -  G^{-}\right]\nonumber\\
\overline{g} &=&  \frac{1}{m} G^{+} + \left(1 - \frac{1}{m}\right)
G^{-} \quad.  
\label{relation1}
\end{eqnarray}

Now we are in a position to calculate correlators $Q_{a b}(s, s'; {\bf 
  k})$ in eq.(\ref{correlator_Q}). The diagonal element $Q_{a a}$ is
\begin{eqnarray}
Q_{a a}(s, s'; {\bf k}) &=& \exp\left\{ - \frac{k^2}{4A}\left[\overline{g}(s,
  s) + \overline{g}(s', s') - 2 \overline{g}(s,
  s')\right]\right\}\nonumber\\
&=&\exp\left\{ - \frac{k^2}{4A m \sinh (\lambda_{+})} \left[ 1 - e^{-
      \lambda_{+}|s - s'|}\right]  - \frac{k^2(m-1)}{4A m \sinh (\lambda_{-})} \left[ 1 - e^{-
      \lambda_{-}|s - s'|}\right] \right\} \quad,
\label{Q_diagonal}
\end{eqnarray}
where we have used eqs. (\ref{relation1}) and (\ref{GF_simple}). The
off - diagonal elements of $Q_{a b}$ inside the blocks are
\begin{eqnarray}
Q^{<}_{a b}(s, s'; {\bf k}) &=& \exp\left\{ -
  \frac{k^2}{4A}\left[\overline{g}(s,s) + \overline{g}(s', s') - 2
    g(s, s')\right]\right\}\nonumber\\
&=&\exp\left\{ - \frac{k^2}{4A m \sinh (\lambda_{+})} \left[ 1 - e^{-
      \lambda_{+}|s - s'|}\right]  - \frac{k^2}{4A m \sinh
    (\lambda_{-})} \left[ m - 1 + e^{-
      \lambda_{-}|s - s'|}\right] \right\} \quad.
\label{Q_off_diagonal}
\end{eqnarray}
Finally the off - diagonal elements of  $Q_{a b}$ outside the blocks can
be written as 
\begin{eqnarray}
Q^{>}_{a b}(s, s'; {\bf k}) &=& \exp\left\{ -
  \frac{k^2}{4A}\left[\overline{g}(s,s) + \overline{g}(s',
    s')\right]\right\}\nonumber\\
&=&\exp\left\{ - \frac{k^2}{4A}\left[ \frac{1}{ m \sinh (\lambda_{+})} 
    + \frac{m - 1}{ m \sinh (\lambda_{-})}\right]\right\} \quad.
\label{Q_outside}
\end{eqnarray}
How many elements $Q_{a b}$  of each type exist? It is easily seen
that the number of diagonal elements is $n$, the number of elements
$Q^{<}_{a b}$ is $n(m - 1)$ and finally the number of elements
$Q^{>}_{a b}$ is $n(n - m)$. After that the interaction part of free
energy (see eq.(\ref{interaction1}) ) can be represented in the
general form
\begin{eqnarray}
F_1 (n, m) &=& - \frac{\Delta}{2} \sum_{s, s'=1}^{N} \int\frac{d^d k}{(2 
  \pi)^d} \biggl\{n \exp\left\{ - \frac{k^2}{4A m \sinh (\lambda_{+})} \left[ 1 - e^{-
      \lambda_{+}|s - s'|}\right]  - \frac{k^2(m-1)}{4A m \sinh (\lambda_{-})} \left[ 1 - e^{-
      \lambda_{-}|s - s'|}\right] \right\} \nonumber\\
&+& n(m - 1)\exp\left\{ - \frac{k^2}{4A m \sinh (\lambda_{+})} \left[ 1 - e^{-
      \lambda_{+}|s - s'|}\right]  - \frac{k^2}{4A m \sinh
    (\lambda_{-})} \left[ m - 1 + e^{-
      \lambda_{-}|s - s'|}\right] \right\} \nonumber\\
&+& n(n - m)\exp\left\{ - \frac{k^2}{4A}\left[ \frac{1}{ m \sinh (\lambda_{+})} 
    + \frac{m - 1}{ m \sinh (\lambda_{-})}\right]\right\} \biggr\}
\label{interaction_general_1}
\end{eqnarray}
or after integration over ${\bf k}$ it can be written as
\begin{eqnarray}
F_1 (n, m) &=&  \frac{\Delta}{2 \pi^{d/2} b^d} \sum_{s, s'=1}^{N}
\biggl\{ n \left[\frac{1 - e^{-\lambda_{+}|s - s'|}}{m \sinh
    (\lambda_{+})} + \frac{m - 1}{m} \:\: \frac{ 1 - e^{- \lambda_{-}|s -
      s'|}}{\sinh(\lambda_{-})}\right]^{- d/2}\nonumber\\
&+& n(m - 1)\left[\frac{1 - e^{-\lambda_{+}|s - s'|}}{m \sinh(\lambda_{+})} + 
\frac{m -  1 + e^{- \lambda_{-}|s -
    s'|}}{m \sinh(\lambda_{-})}\right]^{- d/2}\nonumber\\
&+& n(n - m)  \left[\frac{1}{m \sinh(\lambda_{+})} + \frac{m - 1}{m}
  \:\: \frac{1}{\sinh(\lambda_{-})}\right]^{- d/2} \biggr\}
\label{interaction_general}
\end{eqnarray}

\section{Regularization by the coarse - graining}

First of all it is pertinent to note that the chain statistics is
Gaussian on the interval $0 < s < 1/\lambda_{+}$ (see
eq.(\ref{mean_square})) whereas the collapse shows up on the larger
contour variable, $1/\lambda_{+} < s < N$. The whole collapsed chain
can be seen as made  up of the statistically independent segments of
length $l \approx 1/\lambda_{+}$. The number of such segments $n_l =
N/l = N\lambda_{+} \gg 1$. The statistical independence and the
equipartition of energy imply that the free energy is proportional to
$k_B T n_l$, i.e.
\begin{eqnarray}
F \approx k_B T n_l =  k_B T N \sqrt{\frac{B}{A}} =  k_B T N
\left(\frac{\Delta}{b^{d}}\right)^{\frac{2}{4 - d}}
\label{coarse_graining}
\end{eqnarray}
and we go back to eq.(\ref{minimum}). Now we can consider the chain as 
a string of totally collapsed statistically independent segments of
length $l \approx 1/\lambda_{+}$. This coarse - grained picture
provides a basis for the regularization of the integral in eq. (\ref{RS}).

The reason for the divergency of this integral at small $s$ is the
following. There is a mismatch between the Gaussian chain in the
external field $B$ and the variational method where the value of $B$
is expected to find from the variational free energy minimization. The 
divergency on the interval,  $0 < s < 1/\lambda_{+}$ shows that the
effective segment - segment interaction within the variational method
comes in a contradiction with the Gaussian statistics on this
interval. One way to resolve  this contradiction is to take the
trial free energy in a more general form, $\sum_{s=1}^{N} {\bf R}(s)
B(s - s'){\bf R}(s')$, i.e. now $B(s)$ is a trial function but not a
constant field. That actually what we have considered in the recent
papers \cite{Miglior1}.

Nevertheless, at the moment it is enough to stay with only one
variational parameter $B$ which correctly describes the behavior of
the string of statistically independent segments. In other words,
the variational method should be applied to the coarse - grained model 
which we have discussed above.

Mathematically this simply amounts to the limiting of the contour
variable $s$ on the interval $1/\lambda_{+} < s < \infty$,
i.e. eq.(\ref{RS}) reads
\begin{eqnarray}
F_{\rm RS}^{(1)} = - \frac{\Delta}{2\pi^{d/2} b^d} N 
\left[\sinh(\lambda_{+})\right]^{d/2} \int\limits_{1/\lambda_{+}}^{\infty} ds
\: \left[\frac{1}{\left[1 - e^{- \lambda_{+} s}\right]^{d/2}} -
  1\right] \quad.
\label{regular}
\end{eqnarray}
After that the RS - free energy becomes
\begin{eqnarray}
F_{\rm RS}^{(1)} = - \frac{\Delta}{2\pi^{d/2} b^d} N 
\frac{\left[\sinh(\lambda_{+})\right]^{d/2}}{\lambda_{+}} \:\: I_d \quad,
\label{RS_1}
\end{eqnarray}
where
\begin{eqnarray}
I_d = \int\limits_{1}^{\infty} \: dx  \left[\frac{1}{\left[1 - e^{-
        x}\right]^{d/2}} - 1\right] \quad,
\label{I_d}
\end{eqnarray}
so that the  integral in eq.(\ref{RS}) is regularized.

\section{RSB - free energy functional: limits and interpolation}

Let us calculate first the entropy term (\ref{entropy1}) in the limits
(\ref{limit_gg}) and (\ref{limit_ll}). For the eigenvalue
$\lambda_{-}$ (see eq.(\ref{eigenvalue})) one can write
\begin{eqnarray}
\lambda_{-} = 2\ln \left[\frac{1}{2} \sqrt{\frac{B + 2mC}{A}} + 
\sqrt{1 + \frac{B + 2mC}{4A}}\right] \quad,
\label{lambda_minus}
\end{eqnarray}
i.e. at $C = 0 \quad \lambda_{-} = \lambda_{+}$. We expand
eqs.(\ref{lambda_minus}) with respect to small $C$ retaining only the
main terms. Then the corresponding RSB - entropy term ( after
subtraction of RS - term, i.e. $F_{\rm RSB}^{(0)} \simeq \left[\lim_{n \to 0} 
F_0/n\right] - F_{\rm RS}^{(0)}$) reads
\begin{eqnarray}
F_{\rm RSB}^{(0)} \simeq \frac{m - 1}{m} \: \frac{d N}{2} \:\frac{m C}{4 A} 
\quad.
\label{entropy_small}
\end{eqnarray}
For the large $C$ using again eq.(\ref{lambda_minus}) in
eq.(\ref{entropy1}) yields the result
\begin{eqnarray}
F_{\rm RSB}^{(0)} \simeq \frac{m - 1}{m} \frac{d N}{2}\left[ \ln
\left(\frac{2mC}{A}\right) - 1\right] \quad.
\label{entropy_large}
\end{eqnarray}
The suitable interpolating expression which embraces both,
eqs.(\ref{entropy_small}) and (\ref{entropy_large}) can be written in
the simple form
\begin{eqnarray}
F_{\rm RSB}^{(0)} \simeq \frac{m - 1}{m} \frac{d N}{2}  \ln \left[\frac{2mC}{A}
 + 1 \right] \quad.
\label{entropy_interpol}
\end{eqnarray}
In such form the entropy term has been obtained in ref
\cite{Takada1,Takada2}.

Now we can make the same calculations for $F_1(n,m)$ in
eq.(\ref{interaction_general}). Because the dependence from $C$ is
contained now only in $\lambda_{-}$ it is convenient for small $C$ to represent
\begin{eqnarray}
\lambda_{-} \simeq \lambda_{+}\left(1 + \frac{mC}{B}\right) \quad.
\label{lambda_more}
\end{eqnarray}
Expansion of the first term in eq.(\ref{interaction_general}) has the
form
\begin{eqnarray}
\left[\frac{1 - e^{-\lambda_{+}|s - s'|}}{m \sinh(\lambda_{+})} +
  \frac{m - 1}{m} \:\: \frac{ 1 - e^{- \lambda_{-}|s -
      s'|}}{\sinh(\lambda_{-})}\right]^{- d/2} &\simeq& \left[\frac{1 -
    e^{-\lambda_{+}|s - s'|}}{\sinh(\lambda_{+})}\right]^{- d/2} +
\frac{d}{2} \: \frac{m - 1}{m} \frac{m C}{B} \left[\frac{1 -
    e^{-\lambda_{+}|s - s'|}}{m \sinh(\lambda_{+})}\right]^{-
  d/2}\nonumber\\
&\times& \left[\frac{\lambda_{+}|s - s'| e^{-\lambda_{+}|s - s'|}}{1 -
    e^{-\lambda_{+}|s - s'|}} - \lambda_{+}\coth(\lambda_{+})\right] \quad.
\label{expansion_first}
\end{eqnarray}
By the same way the 2nd and 3rd terms in
eq.(\ref{interaction_general}) yields correspondingly
\begin{eqnarray}
\left[\frac{1 - e^{-\lambda_{+}|s - s'|}}{m \sinh(\lambda_{+})} + 
\frac{m -  1 + e^{- \lambda_{-}|s -
    s'|}}{\sinh(\lambda_{-})}\right]^{- d/2} &\simeq&
\left[\sinh(\lambda_{+})\right]^{d/2} + 
\frac{d}{2} \left[\sinh(\lambda_{+})\right]^{d/2}
\frac{mC}{B}\nonumber\\
&\times& \left[\frac{\lambda_{+}|s - s'| e^{-\lambda_{+}|s - s'|}}{m}
  + \frac{\lambda_{+}\coth(\lambda_{+})}{m}\left[m - 1 +
    e^{-\lambda_{+}|s - s'|}\right] \right]
\label{expansion_second}
\end{eqnarray}
and
\begin{eqnarray}
\left[\frac{1}{m \sinh(\lambda_{+})} + \frac{m - 1}{m}
  \:\: \frac{1}{\sinh(\lambda_{-})}\right]^{- d/2} \simeq
\left[\sinh(\lambda_{+})\right]^{d/2} + \frac{d}{2} \:  \frac{m - 1}{m}
\frac{m C}{B} \left[\sinh(\lambda_{+})\right]^{d/2}\lambda_{+}\coth(\lambda_{+}) 
\label{expansion_third}
\end{eqnarray}

The 3rd term in eq.(\ref{interaction_general}) has the factor $n(n - m)$ which at $n \to 0$ takes the
form $n(n - m) \approx - n - n(m - 1)$, so that we can combine the 3rd 
term with the 1st and 2nd terms correspondingly. After that by
switching from the summation to the integration over $s$ and $s'$ we
arrive at the following two contributions to the RSB - interaction
free energy part
\begin{eqnarray}
F_{\rm RSB}^{(1)} = \left[\lim_{n \to 0} \frac{F_1(n,m)}{n}\right] - F_{\rm
  RS}^{(1)} = J_1 + J_2 \quad,
\label{RSB_interaction}
\end{eqnarray}
where
\begin{eqnarray}
J_1 &=& - \frac{\Delta d N}{2\pi^{d/2} b^d} \:  \frac{m - 1}{m}
\: \frac{m C}{B} \:
\frac{\left[\sinh(\lambda_{+})\right]^{d/2}}{\lambda_{+}}
\left[\lambda_{+}\coth(\lambda_{+}) - \frac{2}{d}\right]
I_d\nonumber\\
J_2 &=& - \frac{\Delta d N}{2\pi^{d/2} b^d} \:  \frac{m - 1}{m}
\: \frac{m C}{B}\:
\frac{\left[\sinh(\lambda_{+})\right]^{d/2}}{\lambda_{+}}
\left[\lambda_{+}\coth(\lambda_{+}) + 1\right] \quad,
\label{two_contributions}
\end{eqnarray}
where $I_d$ is given by eq.(\ref{I_d}). At small $\lambda_{+}$ it can be
seen  that $\lambda_{+}\coth(\lambda_{+}) \simeq 1 + \lambda_{+}^2/2$
and both terms, $J_1$ and $J_2$ are negative. Then the resulting
expression for $F_{\rm RSB}^{(1)}$ at small $C$ is given by
\begin{eqnarray}
F_{\rm RSB}^{(1)} = - \frac{\Delta d N}{2\pi^{d/2} b^d} \:  \frac{m - 1}{m}
\: \frac{m C}{B} \: \frac{\left[\sinh(\lambda_{+})\right]^{d/2}}{\lambda_{+}}
\label{RSB_interaction_1}
\end{eqnarray}

In the limit of large $C$ the eigenvalue $\lambda_{-}$ is large also and all terms
which include $1/\sinh(\lambda_{-})$ can be neglected. After that from 
eq.(\ref{interaction_general}) for $F_{\rm RSB}^{(1)} = [\lim_{n \to
  0} F_1(n,m)/n] - F_{\rm RS}^{(0)}$ we immediately have
\begin{eqnarray}
F_{\rm RSB}^{(1)} = - \frac{\Delta d N}{2\pi^{d/2} b^d} \: 
\:\left(m^{\frac{d}{2} + 1} - 1\right) \:
\frac{\left[\sinh(\lambda_{+})\right]^{d/2}}{\lambda_{+}} \: I_d
\label{RSB_interaction_2}
\end{eqnarray}
Let us denote
\begin{eqnarray}
y \equiv \frac{2 m C}{A} = 2 m C b^2 \quad.
\label{notation}
\end{eqnarray}
Then the parameter $m C/B \simeq y/(2\lambda_{+}^2)$, where we have
used $\lambda_{+}^2 \simeq B/A$. After that the expression which
interpolates between eqs.(\ref{RSB_interaction_1}) and
(\ref{RSB_interaction_2}) can be written in the form
\begin{eqnarray}
F_{\rm RSB}^{(1)} \simeq  - \frac{\Delta d N}{2\pi^{d/2} b^d}
\:\frac{\left[\sinh(\lambda_{+})\right]^{d/2}}{\lambda_{+}}
\:\:\frac{(m^{\frac{d}{2} + 1} - 1) \:\:  y}{y + \frac{m
    (m^{\frac{d}{2} + 1} - 1)}{m - 1} \:2\lambda_{+}^2}
\label{RSB_interpolation}
\end{eqnarray}
\end{appendix}

\end{document}